\documentclass[pre,aps]{revtex4}
\usepackage{graphicx}

\begin{document}

\title{Low-temperature properties of classical zigzag spin chain near the
ferromagnet-helimagnet transition point}
\author{D.~V.~Dmitriev}
\email{dmitriev@deom.chph.ras.ru}
\author{V.~Ya.~Krivnov}
\affiliation{Joint Institute of Chemical Physics, RAS, Kosygin str.
4, 119334, Moscow, Russia.}
\date{}

\begin{abstract}
Low-temperature thermodynamics of the classical frustrated
ferromagnetic spin chain near the ferromagnet-helimagnet transition
point is studied by means of mapping to the continuum limit. The
calculation of the partition function and spin correlation function
is reduced to quantum problem of a particle in potential well. It is
shown that exactly at the transition point the correlation length
behaves as $T^{-1/3}$ and the magnetic susceptibility diverges as
$T^{-4/3}$ in the low-temperature limit. Corresponding numerical
factors for the correlation length and the susceptibility is
calculated. It is shown that the low-temperature susceptibility in
the helical phase near the transition point has a maximum at some
temperature. Such behavior as well as the location and the magnitude
of the maximum as a function of deviation from the transition point
are in agreement with that observed in several materials described
by the quantum $s=1/2$ version of this model.
\end{abstract}

\maketitle

\section{Introduction}

Lately, there has been considerable interest in low-dimensional spin
models that exhibit frustration \cite{mikeskabook}. A very
interesting class of such systems with unique physical properties is
chain compounds consisting of edge-sharing $CuO_{4}$ units
\cite{Mizuno,Masuda,Hase,Malek,Capogna,Nitzsche}. The frustration in
these compounds arises from the competition of the ferromagnetic (F)
interaction $J_{1}$ of nearest neighbor (NN) spins and the
antiferromagnetic (AF) next-nearest-neighbor (NNN) interaction
$J_{2}$. An appropriate model describing the magnetic properties of
such copper oxides is so called F-AF spin chain model, the
Hamiltonian of which has a form
\begin{equation}
H=J_{1}\sum \mathbf{S}_{n}\cdot\mathbf{S}_{n+1} +J_{2}\sum
\mathbf{S}_{n}\cdot\mathbf{S}_{n+2}  \label{H}
\end{equation}
where $J_{1}<0$ and $J_{2}>0$.

This model is characterized by the frustration parameter $\alpha
=J_{2}/|J_{1}|$. The ground state properties of the quantum $s=1/2$
F-AF chain have been intensively studied last years
\cite{Chubukov,Itoi,Dmitriev,Vekua,Lu,Richter,Kuzian,Kecke,Sudan}.
It is known that the ground state of model (\ref{H}) is
ferromagnetic for $\alpha <1/4$. At $\alpha =1/4$ the quantum phase
transition to the incommensurate singlet phase with helical spin
correlations takes place. Remarkably, this transition point does not
depend on a spin value, including the classical limit $s\to\infty $.

However, the influence of the frustration on low-temperature
thermodynamics is less studied, especially in the vicinity of the
ferromagnet-helimagnet transition point. It is of a particular
importance to study this problem, because edge-sharing cuprates with
$\alpha \simeq 1/4$ (for example, $Li_{2}CuZrO_{4}$,
$Rb_{2}Cu_{2}Mo_{3}O_{12}$) are of special interest \cite{Volkova}.
Unfortunately, at present the low-temperature thermodynamics of
quantum $s=1/2$ model (\ref{H}) at $\alpha \neq 0$ can be studied
only either by using of numerical calculations of finite chains or
by approximate methods. On the other hand, the classical version of
model (\ref{H}) can be studied by analytical methods giving exact
results at $T\to 0$. Of course, the question arises about the
relation of these results (in particular, for the susceptibility) to
those of the quantum model. It is known
\cite{Nakamura,Sachdev,Bacalis} that quantum and classical
ferromagnetic chains ($\alpha =0$) have universal low-temperature
behavior. As was noted in Ref.\cite{Sachdev} the physical reason of
this universality is the consequence of the fact that the
correlation length at $T\to 0$ is larger than de Broglie wavelength
of the spin waves. This property is inherent in the frustrated
ferromagnet too. Though such universality for the frustrated
ferromagnetic chains is not strictly checked at present, one can
expect that the universality holds on for the F-AF chain as well.
Therefore, the study of classical model (\ref{H}) can be useful for
the understanding of the low-temperature properties of the quantum
F-AF chains.

At zero temperature classical model (\ref{H}) has long range-order
(LRO) for all values of $\alpha $: the ferromagnetic LRO at
$\alpha\leq 1/4$ and the helical one at $\alpha>1/4$. At finite
temperature the LRO is destroyed by thermal fluctuations and
thermodynamic quantities have singular behavior at $T\to 0$. In
particular, the zero-field magnetic susceptibility $\chi $ diverges.
For the 1D Heisenberg ferromagnet (HF) $\chi=2|J_{1}|/3T^{2}$
\cite{Fisher}. At $0<\alpha <1/4$ the susceptibility is $\chi
=2(1-4\alpha )\left\vert J_{1}\right\vert /3T^{2}$. This behavior of
$\chi$ is similar to that for the quantum $s=1/2$ F-AF model
\cite{Hartel}. The value $\chi T^2$ vanishes at the transition
point. As it was noted in Ref.\cite{Hartel} this fact indicates the
change in the critical exponent.

In this paper we focus on the low-temperature behavior of the
classical F-AF chain near the ferromagnet-helimagnet transition
point. At first we consider the case $\alpha=1/4$, i.e. the F-AF
model exactly at the transition point. This problem is interesting
on its own account, because the spectrum of low-energy excitations
is proportional to $k^4$ rather than $k^2$ as for the HF model. It
means that the critical exponents characterizing the low-temperature
behavior of thermodynamic quantities at $\alpha=1/4$ can be
different from those for the HF chain. Besides, the method developed
for the study of the transition point can be generalized to
investigate the vicinity of the transition point.

At the ferromagnet-helimagnet transition point $\alpha =1/4$ it is
convenient to rewrite Hamiltonian (\ref{H}) in the form \cite{DK10}:
\begin{equation}
H=\frac{1}{8}\sum (\mathbf{S}_{n+1}-2\mathbf{S}_{n}+\mathbf{S}_{n-1})^{2}
\label{H14}
\end{equation}

In Eq.(\ref{H14}) we put $|J_{1}|=1$ and omit unessential constant.

In the classical approximation the spin operators $\mathbf{S}_{n}$
are replaced by the classical vectors $\vec{S}_{n}$ of the unit
length. In what follows we use the continuum approach replacing
$\vec{S}_{n}$ by the classical vector field $\vec{s}(x)$ with slowly
varying orientations, so that
\begin{equation}
\vec{S}_{n+1}-2\vec{S}_{n}+\vec{S}_{n-1}\approx \frac{\partial
^{2}\vec{s}(x_{n})}{\partial x^{2}}  \label{dSdx2}
\end{equation}
where the lattice constant is chosen as unit length.

In the low-temperature limit the thermal fluctuations are weak, so
that neighbor spins are directed almost parallel and continuum
approach (\ref{dSdx2}) is justified. Using the continuum
approximation, Hamiltonian (\ref{H14}) goes over into the energy
functional of the vector field $\vec{s}(x)$:
\begin{equation}
E=\frac{1}{8}\int \mathrm{d}x\left( \frac{\partial ^{2}\vec{s}}{\partial
x^{2}}\right) ^{2}  \label{Etr}
\end{equation}

This energy functional is a starting point of the investigations
of model (\ref{H}) at $\alpha=1/4$. The partition function is a
functional integral over all configurations of the vector field on
a ring of length $L$
\begin{equation}
Z=\int D\vec{s}(x)\exp \left\{ -\frac{1}{8T}\int_{0}^{L}dx\left( \frac{d^{2}%
\vec{s}}{dx^{2}}\right) ^{2}\right\}  \label{Z0}
\end{equation}

It is useful to scale the spatial variable as
\begin{equation}
\xi =2T^{1/3}x  \label{xi}
\end{equation}

Then, the partition function takes the dimensionless form
\begin{equation}
Z=\int D\vec{s}(\xi )\exp \left\{ -\int_{0}^{\lambda }d\xi \left( \frac{d^{2}%
\vec{s}}{d\xi ^{2}}\right) ^{2}\right\}  \label{Zxi}
\end{equation}
where the rescaled system length is $\lambda =2T^{1/3}L$. The
partition function (\ref{Zxi}) and the correlation function
$\left\langle \vec{s}(l)\cdot \vec{s}(0)\right\rangle $ are the
objects of the current study.

The paper is organized as follows. In Sec.II we consider the planar
version of spin model (\ref{Etr}) at $\alpha=1/4$. For this more
simple model we demonstrate the technique of the calculation of the
correlation function. We show that the thermodynamics of this
classical one-dimensional model reduces to the zero-dimensional
quantum problem of a particle in a potential well. In Sec.III the
classical continuum F-AF model at the transition point is studied.
In this case the partition function describes a quantum particle in
an axially symmetrical potential well. We obtain the exact
expressions for the susceptibility and the structure factor. In
Sec.IV the behavior of the uniform susceptibility in the helical
phase at $\alpha\gtrsim 1/4$ is studied and compared with the
experimental data for the edge-shared compounds and with the results
for the quantum $s=1/2$ model. The conclusions are summarized in
Sec.V. and the Appendix contains some technical aspects of the
calculation of the correlation function.

\section{Planar spin case at $\alpha=1/4$}

\subsection{Partition function}

We begin our investigation of the thermodynamics in the transition
point with a more simple planar spin version of model (\ref{Etr}),
when all spin vectors lie in one plane and have only two components:
\begin{equation}
\vec{s}(\xi )=(\sin \theta (\xi ),\cos \theta (\xi ))  \label{s2d}
\end{equation}

Such order of study is methodically justified, because the technique
of the correlation function calculation is similar for both planar
and original three-component spin models, but it is easier to
demonstrate on the simple planar model.

In terms of $\theta (\xi )$ the Hamiltonian transforms to
\begin{equation}
\left( \frac{d^{2}\vec{s}}{d\xi ^{2}}\right) ^{2}=\left( \frac{d^{2}\theta }{%
d\xi ^{2}}\right) ^{2}+\left( \frac{d\theta }{d\xi }\right) ^{4}
\label{stheta}
\end{equation}
and the partition function becomes
\begin{equation}
Z=\int D\theta
(\xi )\exp \left\{ -\int_{0}^{\lambda }d\xi \left( \theta ^{\prime
\prime 2}+\theta ^{\prime 4}\right) \right\}  \label{Ztheta}
\end{equation}
where the prime denotes the space derivatives $d/d\xi $.

In general, when one deals with the field theory containing the second or
higher order derivatives one has to follow the Ostrogradski prescription
\cite{Kleinert}. However, as will be demonstrated below, in our case we can
avoid such complications and calculate the partition function and
correlation functions in a more simple way.

Since the Hamiltonian contains only derivatives of the field
$\theta(\xi)$, the partition function can be rewritten in terms of a
new field
\begin{equation}
Z=\int Dq(\xi )\exp \left\{ -\int_{0}^{\lambda }d\xi \left( q^{\prime
2}+q^{4}\right) \right\}  \label{zq2d}
\end{equation}
where
\begin{equation}
q(\xi )=\frac{d\theta (\xi )}{d\xi }  \label{q2d}
\end{equation}

To calculate the partition function we utilize well-known
equivalence of the $n$-dimensional statistical field theory with the
$(n-1)$-dimensional quantum field theory. It is obvious in advance
that partition function (\ref{zq2d}) describes a quantum particle in
a potential well $U(q)=q^{4}$ at `temperature' $1/\lambda $.
However, we will follow all intermediate steps, because we will need
them in the subsequent calculations of the correlation function.

The transition amplitude (or propagator) of a particle located
initially at $q(0)=q_{i}$, and finally at $q(t)=q_{f}$ takes the
form of a path integral
\begin{equation}
\left\langle q_{f}\right\vert e^{-it\hat{H}}\left\vert q_{i}\right\rangle
\propto \int_{q_{i}}^{q_{f}}Dq(t)\exp \left\{ i\int_{0}^{t}dtL(\dot{q}%
,q)\right\}  \label{propag}
\end{equation}

Then, imposing the periodic boundary conditions $q_{f}=q_{i}=q$ and
integrating over $q$, we obtain the partition function in a form
\begin{equation}
Z\propto \int dq\left\langle q\right\vert e^{-it\hat{H}}\left\vert
q\right\rangle  \label{Zpath}
\end{equation}

In our case we replace $\xi $ by an imaginary time $\xi \to it$ and
partition function (\ref{zq2d}) takes the form of a path integral of
a quantum particle in a potential well:
\begin{equation}
Z=\int Dq(t)\exp \left\{ i\int_{0}^{-i\lambda }dtL_{0}(\dot{q},q)\right\}
\label{ZL02d}
\end{equation}
where the Lagrangian is
\begin{equation}
L_{0}=\dot{q}^{2}-q^{4}  \label{L02d}
\end{equation}

The momentum $p$ is $p=2\dot{q}$ and the Hamiltonian is
\begin{equation}
H_{0}=\frac{1}{4}p^{2}+q^{4}  \label{H02d}
\end{equation}

The corresponding Schr\"{o}dinger equation describes a quantum anharmonic
oscillator:
\begin{equation}
-\frac{1}{4}\frac{d^{2}\psi }{dq^{2}}+q^{4}\psi =\varepsilon \psi
\label{schr2d}
\end{equation}

The spectrum of equation (\ref{schr2d}) is calculated numerically:
\begin{equation}
\varepsilon _{\alpha }=0.4208;1.508;2.96\ldots  \label{e2d}
\end{equation}
and the wave function $\psi _{0}(q)$ for the lowest level $\varepsilon
_{0}=0.4208$ is shown in Fig.1.

\begin{figure}[tbp]
\includegraphics[width=3in,angle=-90]{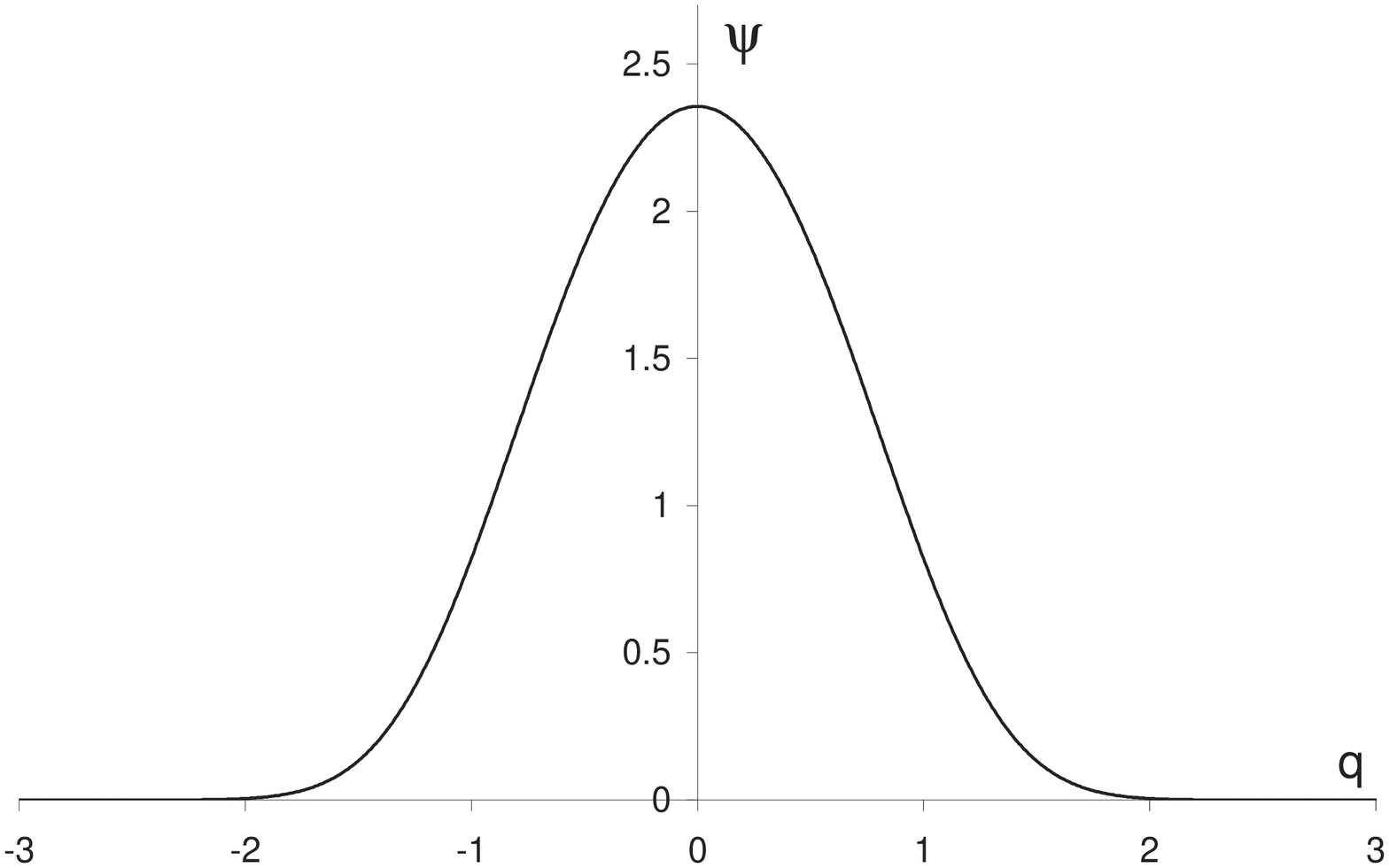}
\caption{Wave function $\psi_0(q)$ for the lowest level
$\varepsilon_0=0.4208$ of planar spin model.}
\end{figure}

Now the exponent of $\hat{H}_{0}$ can be represented as follows:
\begin{equation}
e^{-\lambda \hat{H}_{0}}=\sum_{\alpha }\left\vert \psi _{\alpha
}\right\rangle e^{-\lambda \varepsilon _{\alpha }}\left\langle \psi _{\alpha
}\right\vert  \label{expH02d}
\end{equation}
and the partition function becomes
\begin{equation}
Z\propto \int dq\left\langle q\right\vert e^{-\lambda \hat{H}}\left\vert
q\right\rangle =\sum_{\alpha }e^{-\lambda \varepsilon _{\alpha }}
\label{z2d}
\end{equation}

As expected, we obtain the partition function of a quantum
anharmonic oscillator at `temperature' $1/\lambda$. In the
thermodynamic limit $\lambda=(2T^{1/3}L)\to\infty$ only the lowest
eigenvalue $\varepsilon_0=0.4208$ gives contribution to the
partition function,
\begin{equation}
Z\rightarrow e^{-\lambda \varepsilon_0}  \label{zasymp2d}
\end{equation}

\subsection{Correlation function}

The scalar product of vector fields located on distance $l$ can be written
as
\begin{equation}
\vec{s}(l)\cdot \vec{s}(0)=\cos \left[ \theta (l)-\theta (0)\right] =\cos %
\left[ \int_{0}^{l}\theta ^{\prime }(x)dx\right] =\Re \left[ \exp \left(
i\int_{0}^{\mu }q(\xi )d\xi \right) \right]  \label{s0sl2d}
\end{equation}
where $\mu =2T^{1/3}l$.

Then, the correlation function can be represented as a ratio of two
functional integrals:
\begin{equation}
\left\langle \vec{s}(l)\cdot \vec{s}(0)\right\rangle =\frac{1}{Z}\Re \left[
Z_{c}\right]  \label{corr2d}
\end{equation}%
where denominator $Z$ is already calculated (see Eq.(\ref{z2d})) and
\begin{equation}
Z_{c}=\int Dq\exp \left\{ -\int_{0}^{\lambda }d\xi \left( q^{\prime
2}+q^{4}\right) +i\int_{0}^{\mu }qd\xi \right\}  \label{zc}
\end{equation}

The latter path integral is clearly divided on two parts $[0,\mu ]$
and $[\mu ,\lambda ]$ and $Z_{c}$ can be represented as
\begin{equation}
Z_{c}=\int dq_{0}dq_{\mu }Z_{1}(q_{0},q_{\mu })Z_{2}(q_{\mu },q_{0})
\label{zcz1z2}
\end{equation}
where the propagators $Z_{1}$ and $Z_{2}$ are
\begin{eqnarray}
Z_{1}(q_{0},q_{\mu }) &=&\int_{q_{0}}^{q_{\mu }}Dq\exp \left(
-\int_{0}^{-i\mu }dtL_{1}(\dot{q},q)\right)  \label{z1} \\
Z_{2}(q_{\mu },q_{\lambda }) &=&\int_{q_{\mu }}^{q_{\lambda }}Dq\exp \left(
-\int_{-i\mu }^{-i\lambda }dtL_{0}(\dot{q},q)\right)  \label{z2}
\end{eqnarray}
and periodic boundary condition $q_{\lambda}=q_{0}$ is applied.

The Lagrangian $L_{0}$ is given by Eq.(\ref{L02d}) and
\begin{equation}
L_{1}=\dot{q}^{2}-q^{4}+iq  \label{L1}
\end{equation}

The propagator $Z_{2}$ is calculated straightforward using
Eqs.(\ref{propag}) and (\ref{expH02d}):
\begin{equation}
Z_{2}=\sum_{\alpha }e^{-(\lambda -\mu )\varepsilon _{\alpha }}\left\langle
q_{\mu }|\psi _{\alpha }\right\rangle \left\langle \psi _{\alpha
}|q_{0}\right\rangle  \label{z2eq}
\end{equation}

But the propagator $Z_{1}$ requires special treatment, because
$L_{1}$ and the corresponding quantum Hamiltonian $H_{1}$ are
non-Hermitian:
\begin{equation}
\hat{H}_{1}=-\frac{1}{4}\frac{d^{2}}{dq^{2}}+q^{4}-iq  \label{H1q2d}
\end{equation}

Non-Hermitian operator $\hat{H}_{1}$ can be represented as
\begin{equation}
\hat{H}_{1}=\sum_{\alpha }\eta _{\alpha }\left\vert u_{\alpha }\right\rangle
\left\langle v_{\alpha }\right\vert  \label{H1uv2d}
\end{equation}
where $\left\vert u_{\alpha }\right\rangle $ and $\left\vert v_{\alpha
}\right\rangle $ are eigenfunctions of direct and conjugate eigenvalue
equations:
\begin{eqnarray}
\hat{H}_{1}\left\vert u_{\alpha }\right\rangle &=&\eta _{\alpha }\left\vert
u_{\alpha }\right\rangle  \nonumber \\
\hat{H}_{1}^{\dagger }\left\vert v_{\alpha }\right\rangle &=&\eta
_{\alpha }^{\ast }\left\vert v_{\alpha }\right\rangle
\label{H1eigen2d}
\end{eqnarray}

The normalization conditions for $\left\vert u_{\alpha
}\right\rangle $ and $\left\vert v_{\alpha }\right\rangle$ are
\begin{equation}
\left\langle v_{\alpha }|u_{\beta }\right\rangle =\left\langle u_{\alpha
}|v_{\beta }\right\rangle =\delta _{\alpha ,\beta }  \label{n2d}
\end{equation}

Equations (\ref{H1eigen2d}) for Hamiltonian (\ref{H1q2d}) transform
to each other by complex conjugation operation. This implies that
eigenfunctions in Eq.(\ref{H1eigen2d}) satisfy the relation
$\left\vert v_{\alpha }\right\rangle =\left\vert u_{\alpha }^{\ast
}\right\rangle$. Thus, we need to solve only one of the differential
equations (\ref{H1eigen2d}). The numerical calculations show that
all eigenvalues of Eq.(\ref{H1eigen2d}) are real and positive. A few
lowest eigenvalues are presented in Eq.(\ref{eta2d}),
\begin{equation}
\eta_{\alpha}=0.6472;1.517;2.99\ldots  \label{eta2d}
\end{equation}

Real and imaginary parts of $u_{0}(q)$ for the lowest level
$\eta_{0}=0.6472$ are shown in Fig.2.

\begin{figure}[tbp]
\includegraphics[width=3in,angle=-90]{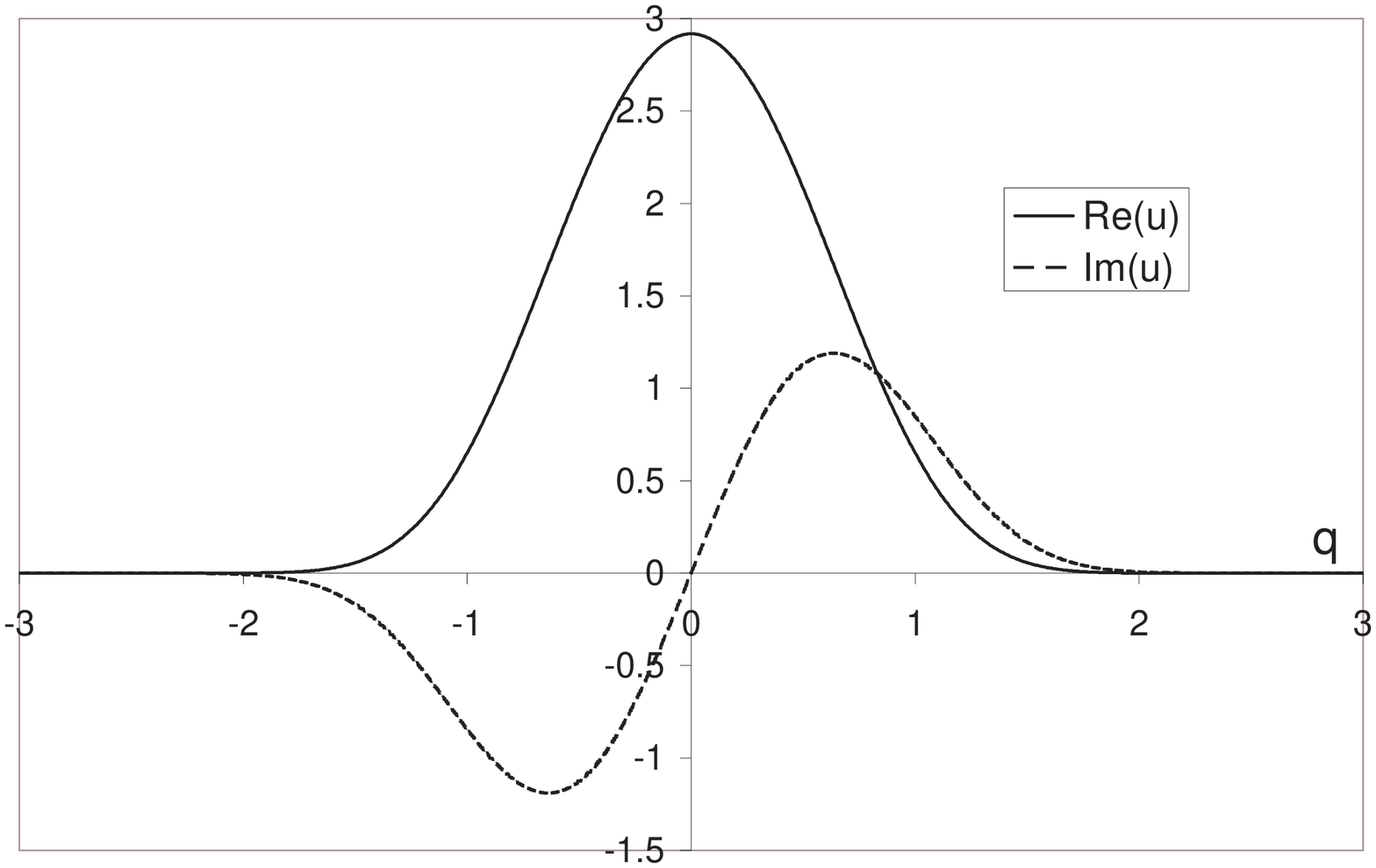}
\caption{Real and imaginary parts of the wave function $u_{0}(q)$
for the lowest level $\eta_0=0.6472$ of planar spin model.}
\label{Rephi}
\end{figure}

Now, the propagator $Z_{1}$ can be expressed through the solutions
of Eq.(\ref{H1eigen2d}),
\begin{equation}
Z_{1}=\left\langle q_{0}\right\vert e^{-\mu \hat{H}_{1}}\left\vert q_{\mu
}\right\rangle =\sum_{\alpha }e^{-\mu \eta _{\alpha }}\left\langle
q_{0}|u_{\alpha }\right\rangle \left\langle v_{\alpha }|q_{\mu }\right\rangle
\label{z1eq}
\end{equation}
where we used the identity
\begin{equation}
e^{-\mu \hat{H}_{1}}=\sum_{\alpha }\left\vert u_{\alpha }\right\rangle
e^{-\mu \eta _{\alpha }}\left\langle v_{\alpha }\right\vert  \label{expH12d}
\end{equation}

Then, substituting Eqs.(\ref{z2eq}) and (\ref{z1eq}) into
Eq.(\ref{zcz1z2}) and integrating over $q_{0},q_{\mu }$ we obtain
\begin{equation}
Z_{c}=\sum_{\alpha ,\beta }e^{-\mu (\eta _{\alpha }-\varepsilon _{\beta
})}e^{-\lambda \varepsilon _{\beta }}\left\langle \psi _{\beta }|u_{\alpha
}\right\rangle ^{2}  \label{zceq}
\end{equation}

The main contribution in the thermodynamic limit $\lambda \to \infty
$ is given by the lowest value $\varepsilon _{0}$ and $Z_{c}$
reduces to
\begin{equation}
Z_{c}\rightarrow e^{-\lambda \varepsilon _{0}}\sum_{\alpha }e^{-\mu (\eta
_{\alpha }-\varepsilon _{0})}\left\langle \psi _{0}|u_{\alpha }\right\rangle
^{2}  \label{zceq2}
\end{equation}

Substituting $\mu =2T^{1/3}l$ into Eq.(\ref{zceq2}), we find from
Eq.(\ref{corr2d}) the correlation function
\begin{equation}
\left\langle \vec{s}(l)\cdot \vec{s}(0)\right\rangle =\Re \sum_{\alpha
}\left\langle \psi _{0}|u_{\alpha }\right\rangle ^{2}e^{-2T^{1/3}(\eta
_{\alpha }-\varepsilon _{0})l}  \label{correq2d}
\end{equation}

The correlation length is governed by the lowest eigenvalue $\eta
_{0}$ and equals to
\begin{equation}
l_{c}=\frac{1}{2T^{1/3}(\eta _{0}-\varepsilon _{0})}=2.2T^{-1/3}
\label{lc2d}
\end{equation}

So, the low-temperature behavior of the correlation length is
different from the HF model, where $l_{c}\sim T^{-1}$.

Now, the structure factor can be also calculated
\begin{equation}
S(k)=2\Re \int_{0}^{\infty }dle^{ikl}\left\langle \vec{s}(l)\cdot \vec{s}%
(0)\right\rangle =2\sum_{\alpha }\left\langle \psi _{0}|u_{\alpha
}\right\rangle ^{2}\frac{2T^{1/3}(\eta _{\alpha }-\varepsilon _{0})}{%
4T^{2/3}(\eta _{\alpha }-\varepsilon _{0})^{2}+k^{2}}  \label{Sk2d}
\end{equation}

In the low-temperature limit, the expansion of the structure factor
for any $ k\gg T^{1/3}$ has the form
\begin{equation}
S(k)=\frac{4T^{1/3}}{k^{2}}\sum_{\alpha }(\eta _{\alpha }-\varepsilon
_{0})\left\langle \psi _{0}|u_{\alpha }\right\rangle ^{2}-\frac{16T}{k^{4}}%
\sum_{\alpha }\left\langle \psi _{0}|u_{\alpha }\right\rangle ^{2}(\eta
_{\alpha }-\varepsilon _{0})^{3}+\ldots  \label{SkT02d}
\end{equation}

The first term in Eq.(\ref{SkT02d}) is zero, because
\begin{equation}
\sum_{\alpha }(\eta _{\alpha }-\varepsilon _{0})\left\langle \psi
_{0}|u_{\alpha }\right\rangle ^{2}=\left\langle \psi _{0}\right\vert \hat{H}%
_{1}-\hat{H}_{0}\left\vert \psi _{0}\right\rangle =-i\left\langle \psi
_{0}\right\vert q\left\vert \psi _{0}\right\rangle =0  \label{term1}
\end{equation}%
and $\left\vert \psi _{0}\right\rangle $ is even function of $q$.

Therefore, the structure factor is given by the second term in
Eq.(\ref{SkT02d}), which can be calculated exactly
\begin{equation}
\sum_{\alpha }(\eta _{\alpha }-\varepsilon _{0})^{3}\left\langle \psi
_{0}|u_{\alpha }\right\rangle ^{2}=\left\langle \psi _{0}\right\vert (\hat{H}%
_{1}-\hat{H}_{0})^{3}+\frac{1}{2}\left[ \left[ \hat{H}_{0},\hat{H}_{1}\right]
,\hat{H}_{1}\right] \left\vert \psi _{0}\right\rangle =-\frac{1}{4}
\label{term2}
\end{equation}

Therefore, the low-temperature asymptotic of the structure factor is
\begin{equation}
S(k)=\frac{4T}{k^{4}}  \label{Skeq2D}
\end{equation}

Hence, the susceptibility $\chi (k)$ for $k\gg T^{1/3}$ remains
finite in the low-temperature limit:
\begin{equation}
\chi (k)=\frac{S(k)}{2T}=\frac{2}{k^{4}}  \label{chik2d}
\end{equation}

The fact that $\chi (k)\sim k^{-4}$ (instead of $k^{-2}$ for HF
chain) is a consequence of the fact that the excitation spectrum
becomes $\sim k^{4}$ at the transition point.

For $k=0$ the structure factor (\ref{Sk2d}) diverges at $T\to 0$ as
\begin{equation}
S(0)=T^{-1/3}\sum_{\alpha }\frac{\left\langle \psi _{0}|u_{\alpha
}\right\rangle ^{2}}{\eta _{\alpha }-\varepsilon _{0}}  \label{S02d}
\end{equation}

The sum in Eq.(\ref{S02d}) is calculated numerically and gives the
factor $\approx 5.36$. Therefore, the magnetic susceptibility
behaves as $T^{-4/3}$
\begin{equation}
\chi (0)=\frac{S(0)}{2T}=\frac{2.68}{T^{4/3}}  \label{chi02d}
\end{equation}

In conclusion of this section we emphasize that the exact
calculation of the correlation function for the planar spin model
demonstrates that the critical exponents at the transition point of
the F-AF chain can differ from that for the HF chain.

\section{Classical spin model at the transition point}

\subsection{Partition function}

The calculation of the correlation functions for the classical
three-component spin model (\ref{H14}) is to a large extent similar
to the planar spin case and to avoid duplications we will often
refer to the previous section. So, in this section we consider the
continuous model described by energy functional (\ref{Etr}) where
three-component vector field $\vec{s}(\xi )$ has the constraint
$\vec{s}^{2}(\xi )=1$.

Since in the partition function the integration occurs over all
possible spin configurations, we are free to choose any local
coordinate system. It is convenient to choose it so that the $Z$
axis at the point $\xi $ is directed along the spin vector
$\vec{s}(\xi )$, so that the spin vector $\vec{s}(\xi )=(0,0,1)$.

Let us introduce a new vector field
\begin{equation}
\vec{q}(\xi )=\frac{d\vec{s}}{d\xi }=(q_{x},q_{y},q_{z})  \label{q3d}
\end{equation}

The constraint $\vec{s}^{2}(\xi )=1$ converts to the relations for
$\vec{q}(\xi )$:
\begin{eqnarray}
q_{z} &=&0  \nonumber \\
q_{z}^{\prime } &=&-q_{x}^{2}-q_{y}^{2}  \label{constrain}
\end{eqnarray}
where the prime denotes the space derivatives $d/d\xi$.

Then, the Hamilton function in Eq.(\ref{Zxi}) transforms to
\begin{equation}
\left( \frac{d^{2}\vec{s}}{d\xi ^{2}}\right) ^{2}=\left( \frac{d\vec{q}}{%
d\xi }\right) ^{2}=q_{x}^{\prime 2}+q_{y}^{\prime
2}+(q_{x}^{2}+q_{y}^{2})^{2}  \label{sq3d}
\end{equation}

Here we see that the constraint $\vec{s}^{2}=1$ effectively eliminates the Z
component of $\vec{q}$ from the Hamilton function. Therefore, henceforth we
deal with the $q_{x}$ and $q_{y}$ components of the vector field $\vec{q}$
only, and we denote a two-component vector field by $\mathbf{q}(\xi
)=(q_{x},q_{y})$.

The partition function in terms of $\mathbf{q}(\xi )$ takes the form:
\begin{equation}
Z=\int D\mathbf{q}\exp \left\{ -\int_{0}^{\lambda }d\xi \left( \mathbf{q}%
^{\prime 2}+\mathbf{q}^{4}\right) \right\}  \label{zq3d}
\end{equation}

Similar to the planar spin case, we treat the partition function as path
integral (\ref{ZL02d}) for the quantum mechanics of a single particle with
the Hamiltonian
\begin{equation}
\hat{H}_{0}=-\frac{1}{4}\Delta +\mathbf{q}^{4}  \label{H0q3d}
\end{equation}
where $\Delta =\partial _{x}^{2}+\partial _{y}^{2}$ is
two-dimensional Laplace operator. Hamiltonian (\ref{H0q3d}) commutes
with the z-component of the angular momentum $\hat{l}_{z}$ and
eigenstates $\psi(\mathbf{q})$ of the corresponding Schrodinger
equation
\begin{equation}
\hat{H}_{0}\psi =\varepsilon \psi  \label{eigen03d}
\end{equation}
are divided to subspaces of azimuthal quantum numbers $l_{z}=0,\pm 1,\pm
2\ldots $.

Thus, the wave function $\psi _{l_{z}}(\mathbf{q})$ describes a
particle with the azimuthal quantum number $l_{z}$ in 2D axially
symmetrical potential well $U(q)=q^{4}$. Numerical solution of
Eq.(\ref{eigen03d}) gives the lowest levels for $l_{z}=0,\pm 1,\pm
2$
\begin{eqnarray}
\varepsilon _{\alpha }(l_{z} &=&0)=0.9305;3.78;7.44\ldots  \nonumber \\
\varepsilon _{\alpha }(l_{z} &=&\pm 1)=2.14;5.48;9.44\ldots  \nonumber \\
\varepsilon _{\alpha }(l_{z} &=&\pm 2)=3.54;7.27;11.5\ldots  \label{e3d}
\end{eqnarray}

The wave function $\psi _{0}(q)$ for the lowest eigenvalue
$\varepsilon _{0}=0.9305$ is shown in Fig.3.

\begin{figure}[tbp]
\includegraphics[width=3in,angle=-90]{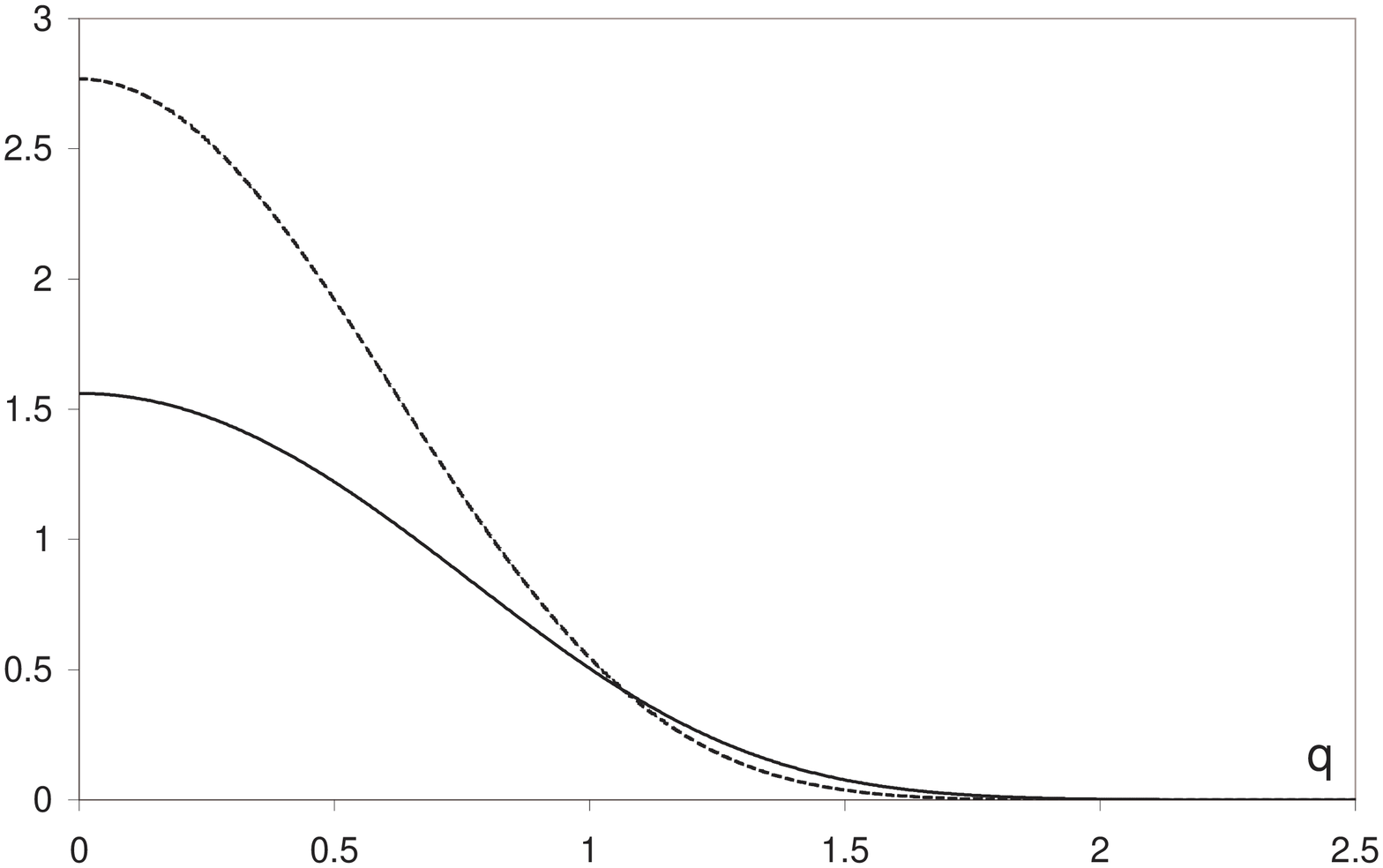}
\caption{Wave functions $\psi(q)$ (solid line) for the lowest
level $\varepsilon_0=0.9305$ and $u_3(q)$ (dashed line) for the
level $\eta_0=1.4113$.} \label{psi}
\end{figure}

Now, by the analogy with the planar spin case
Eqs.(\ref{expH02d})-(\ref{z2d}) we obtain the partition function as
a sum of exponents over all quantum numbers $\alpha $ and $l_{z}$:
\begin{equation}
Z=\sum_{\alpha ,l_{z}}e^{-\lambda \varepsilon _{\alpha ,l_{z}}}  \label{z03d}
\end{equation}

In the thermodynamic limit $\lambda \to \infty $ only the lowest
level $\varepsilon _{0}=0.9305$ survives and the partition function
is
\begin{equation}
Z=e^{-\lambda \varepsilon _{0}}  \label{Z3d}
\end{equation}

\subsection{Correlation function}

Since we work with the local coordinate system directed so that the
spin vector is directed along the $Z$ axis, in order to find the
scalar product of spin vectors $\vec{s}(l)\cdot \vec{s}(0)$ we need
to express the vector $\vec{s}(0)$ in the local coordinate system
located at the point $x=l$. This can be represented as a chain of
successive rotations describing the trajectory $\vec{s}(x)$ [see the
Appendix]:
\begin{equation}
\vec{s}_{x=l}(0)=\exp \left( i\int_{0}^{l}\Omega dx\right)
\vec{s}_{x=0}(0) \label{sxl}
\end{equation}
where $\Omega =\left( \vec{\sigma}\cdot \vec{\omega}(x)\right) $ can
be expressed through the vector $\vec{q}(x)$ as
\begin{equation}
\Omega =\frac{(\vec{\sigma}\cdot \lbrack \vec{q}\times \vec{q}^{\prime }])}{%
\vec{q}^{2}}=\sigma _{y}q_{x}-\sigma _{x}q_{y}+\sigma _{z}\frac{%
q_{x}q_{y}^{\prime }-q_{y}q_{x}^{\prime }}{q_{x}^{2}+q_{y}^{2}}
\label{omega}
\end{equation}

In this equation we used relations (\ref{constrain}) for the $Z$
component of $\vec{q}$ and $\vec{q}^{\prime}$.

Then, taking into account that the spin vectors are directed along
the $Z$ axis of local coordinate system
$\vec{s}_{x=0}(0)=\vec{s}_{x=l}(l)=(0,0,1)$, the scalar product of
spin vectors $\vec{s}(l)\cdot \vec{s}(0)$ becomes:
\begin{equation}
\vec{s}(l)\cdot \vec{s}(0)=\left(
\begin{array}{ccc}
0 & 0 & 1%
\end{array}%
\right) \exp \left( i\int_{0}^{l}\Omega (\mathbf{q},\mathbf{q}^{\prime
})dx\right) \left(
\begin{array}{c}
0 \\
0 \\
1%
\end{array}%
\right)  \label{s0sl3d}
\end{equation}

Similar to the planar case, the correlation function can be written
as a ratio of two path integrals (\ref{corr2d}), where the
denominator $Z$ is given by Eq.(\ref{Z3d}) and the numerator
represents the following path integral:
\begin{equation}
\int D\mathbf{q}\exp \left\{ -\int_{0}^{\lambda }d\xi \left( \mathbf{q}%
^{\prime 2}+\mathbf{q}^{4}\right) +i\int_{0}^{\mu }\Omega (\mathbf{q},%
\mathbf{q}^{\prime })d\xi \right\}  \label{pathint3d}
\end{equation}

Then, repeating the arguments presented in Eqs.(\ref{zc})-(\ref{z2eq}), we
arrive at the problem of the calculation of the propagator:
\begin{equation}
\int_{q_{0}}^{q_{\mu }}D\mathbf{q}\exp \left\{ i\int_{0}^{-i\mu }dtL_{1}(%
\mathbf{\dot{q}},\mathbf{q})\right\} =\left\langle q_{0}\right\vert e^{-\mu
\hat{H}_{1}}\left\vert q_{\mu }\right\rangle  \label{path3d}
\end{equation}
where the Lagrangian and the corresponding quantum Hamiltonian are
\begin{eqnarray}
L_{1} &=&\mathbf{\dot{q}}^{2}-\mathbf{q}^{4}-i\sigma _{y}q_{x}+i\sigma
_{x}q_{y}-\sigma _{z}\frac{q_{x}\dot{q}_{y}-q_{y}\dot{q}_{x}}{%
q_{x}^{2}+q_{y}^{2}}  \label{L13d} \\
\hat{H}_{1} &=&-\frac{1}{4}\Delta +\mathbf{q}^{4}+\frac{\sigma
_{z}^{2}-2\sigma _{z}\hat{l}_{z}}{4\mathbf{q}^{2}}+i\sigma _{y}q_{x}-i\sigma
_{x}q_{y}  \label{H13d}
\end{eqnarray}

Substituting $\sigma _{\alpha }$ from Eqs.(\ref{A7}), the
Hamiltonian takes the matrix form
\begin{equation}
\hat{H}_{1}=\left(
\begin{array}{ccc}
\hat{H}_{0}+\frac{1}{4\mathbf{q}^{2}} & \frac{1}{2\mathbf{q}^{2}}i\hat{l}_{z}
& q_{x} \\
-\frac{1}{2\mathbf{q}^{2}}i\hat{l}_{z} & \hat{H}_{0}+\frac{1}{4\mathbf{q}^{2}%
} & q_{y} \\
-q_{x} & -q_{y} & \hat{H}_{0}%
\end{array}%
\right)  \label{H1matrix}
\end{equation}
where $\hat{H}_{0}$ is defined by Eq.(\ref{H0q3d}).

Operator $\hat{H}_{1}$ is non-Hermitian and the exponent of
$\hat{H}_{1}$ can be represented as
\begin{equation}
e^{-\mu \hat{H}_{1}}=\sum_{\alpha }\left\vert \vec{u}_{\alpha }\right\rangle
e^{-\mu \eta _{\alpha }}\left\langle \vec{v}_{\alpha }\right\vert
\label{expH13d}
\end{equation}%
where three-component eigenvectors $\vec{u}(q_{x},q_{y})=\left(
u_{1},u_{2},u_{3}\right) $ and $\vec{v}(q_{x},q_{y})=\left(
v_{1},v_{2},v_{3}\right) $ satisfy the corresponding eigenvalue
equations
\begin{eqnarray}
\hat{H}_{1}\vec{u} &=&\eta\vec{u}  \nonumber \\
\hat{H}_{1}^{\dagger}\vec{v} &=&\eta^{\ast}\vec{v} \label{eigenH13d}
\end{eqnarray}

The normalization conditions are
\begin{equation}
\left\langle \vec{v}_{\alpha }|\vec{u}_{\beta }\right\rangle =\left\langle
\vec{u}_{\alpha }|\vec{v}_{\beta }\right\rangle =\delta _{\alpha ,\beta }
\label{n3d}
\end{equation}

Making the same procedure for non-Hermitian operators as for the
planar spin case we obtain the correlation function in a form
\begin{equation}
\left\langle \vec{s}(l)\cdot \vec{s}(0)\right\rangle
=\Re\sum_{\alpha} \left\langle \psi_0|u_{3,\alpha }\right\rangle
\left\langle v_{3,\alpha }|\psi_0\right\rangle e^{-\mu(\eta _{\alpha
}-\varepsilon _{0})}  \label{corr3d}
\end{equation}

Only the eigenfunctions $\left\vert u_{3,\alpha }\right\rangle$ and
$\left\vert v_{3,\alpha }\right\rangle $ are present in the above
equation, because according to Eq.(\ref{s0sl3d}) we need only the
element $(3,3)$ of the resultant matrix.

Since the wave function $\psi_0$ has zero angular momentum, then
only the sector $l_{z}=0$ of Eqs.(\ref{eigenH13d}) gives the
contribution to the correlation function. In this sector the wave
functions depend only on $q=|\mathbf{q}|$, the Hamiltonian is
simplified so that we have to solve a pair (instead of three)
differential equations for $u_{3}(q)$ and
$\phi(q)=(q_xu_1+q_yu_2)/q$:
\begin{eqnarray}
-\frac{1}{4}\frac{d^{2}\phi }{dq^{2}}-\frac{1}{4q}\frac{d\phi
}{dq}+\frac{1}{4q^{2}}\phi +q^{4}\phi +qu_{3} &=&\eta \phi  \nonumber \\
-\frac{1}{4}\frac{d^{2}u_{3}}{dq^{2}}-\frac{1}{4q}\frac{du_{3}}{dq}
+q^{4}u_{3}-q\phi &=&\eta u_{3}  \label{ode2}
\end{eqnarray}

The conjugate eigenvalue problem for $\vec{v}$ in
Eq.(\ref{eigenH13d}) transforms to exactly the same differential
equations (\ref{ode2}) for the functions $v^*_{3}(q)$ and
$\chi^*(q)=-(q_xv^*_1+q_yv^*_2)/q$. Therefore, the function $v_3(q)$
is found from the solution of Eq.(\ref{ode2}) by the relation
$v_3(q)=u^*_3(q)$ and the the normalization conditions (\ref{n3d})
transform to
\begin{equation}
\left\langle u^*_{3,\alpha }|u_{3,\beta }\right\rangle -\left\langle
\phi^*_{\alpha }|\phi_{\beta }\right\rangle =\delta _{\alpha ,\beta
} \label{norma3d}
\end{equation}

One can see that equations (\ref{ode2}) describe a two-level system
in an axially symmetric potential well $U(q)=q^{4}$, where two
levels with angular momenta $l_{z}=0$ and $l_{z}=1$ are coupled by
non-Hermitian transition operator. The spectrum of this system of
equations turns out to be real and positive as for the planar case
and a few lowest levels are:
\begin{equation}
\eta _{\alpha }=1.4113;1.83;3.98\ldots  \label{eta3d}
\end{equation}

The reality of the spectrum $\eta_{\alpha}$ has an important
consequence: the correlation function (\ref{corr3d}) decays on large
distances without oscillations.

The wave function $u_{3}(q)$ for the lowest level $\eta _{0}=1.4113$
is shown in Fig.3. As follows from Fig.3 the behavior of the
functions $\psi_{0}(q)$ and $u_{3}(q)$ are similar. Therefore, the
main contribution to the correlation function, the structure factor
and the susceptibility is given by the level $\eta_0$.

The correlation function and the structure factor are given by
equations (\ref{correq2d}), (\ref{Sk2d}) with the substitution
$u_{3,\alpha }$ for $u_{\alpha }$ and eigenvalues presented in
Eqs.(\ref{e3d}), (\ref{eta3d}). Therefore, the correlation length
defined by the lowest eigenvalue $\eta_0$ behaves similar to the
planar spin case $\sim T^{-1/3}$, but the numerical factor is
different:
\begin{equation}
l_{c}=1.04T^{-1/3}  \label{lc3d}
\end{equation}

The low-temperature asymptotic of static structure factor $S(k)$ and
the susceptibility $\chi(k)$ for $k\gg T^{1/3}$ is calculated in a
similar way as for the planar spin case (see Eq.(\ref{SkT02d}))
resulting in
\begin{eqnarray}
S(k) &=&\frac{8T}{k^{4}}  \nonumber \\
\chi (k) &=&\frac{8}{3k^{4}}  \label{Skchik3d}
\end{eqnarray}

For $k=0$ the structure factor $S(0)$ is defined by
Eq.(\ref{S02d}) with the corresponding eigenfunctions and
eigenvalues. The numerical calculation of the sum gives
\begin{equation}
S(0)=T^{-1/3}\sum_{\alpha }\frac{\left\langle \psi _{0}|u_{3,\alpha
}\right\rangle ^{2}}{\eta _{\alpha }-\varepsilon _{0}}=\frac{3.21}{T^{1/3}}
\label{s03d}
\end{equation}

Thus, we have arrived at the final result for the magnetic
susceptibility of the classical spin model at the transition point:
\begin{equation}
\chi (0)=\frac{S(0)}{3T}=\frac{1.07}{T^{4/3}}  \label{chi03d}
\end{equation}

We see that the planar spin model gives correct critical exponent for the
magnetic susceptibility.

\section{Low-temperature susceptibility in the helical phase}

In the preceding sections we have considered the low-temperature
thermodynamics at the transition point $\alpha =1/4$. Likewise it is
possible to extend the developed method for the study of the
vicinity of the transition point. In this case an additional term
$2(\alpha -1/4)\vec{s}^{\prime 2}$ appears in energy functional
(\ref{Etr}), which after rescaling of spatial variable (\ref{xi})
forms the scaling variable
\begin{equation}
\gamma =\frac{\alpha-1/4}{T^{2/3}}  \label{gamma}
\end{equation}

Especially interesting is to study the vicinity of the transition
point when values $(\alpha-1/4)$ and $T$ are small but the
parameter $\gamma$ is fixed. All the steps in derivation of the
expressions for the correlation function are exactly the same as
was done for the transition point and only the form of the
potential well in the corresponding differential equations
(\ref{ode2}) is modified: $U(q)=q^4-4\gamma q^2$. Numerical
solution of the corresponding differential equations allows to
find the correlation function, the static structure factor $S(k)$
and the susceptibility as a function of $\gamma$. As follows from
Eqs.(\ref{s03d}) and (\ref{chi03d}) the susceptibility can be
rewritten as
\begin{equation}
\chi=\frac{f(\gamma)}{T^{4/3}}=\frac{\gamma^2 f(\gamma)}{(\alpha
-1/4)^2} \label{chiT}
\end{equation}

Thus, the normalized susceptibility $\widetilde{\chi }=\chi(\alpha
-1/4)^2$ is a universal function of $\gamma$.

On the ferromagnetic side of the transition point ($\alpha<1/4$)
the susceptibility diverges at $T\to 0$, but the exponent changes
from $4/3$ to $2$, so that the susceptibility becomes
$\chi\sim(1/4-\alpha)/T^2$.

\begin{figure}[tbp]
\includegraphics[width=3in,angle=-90]{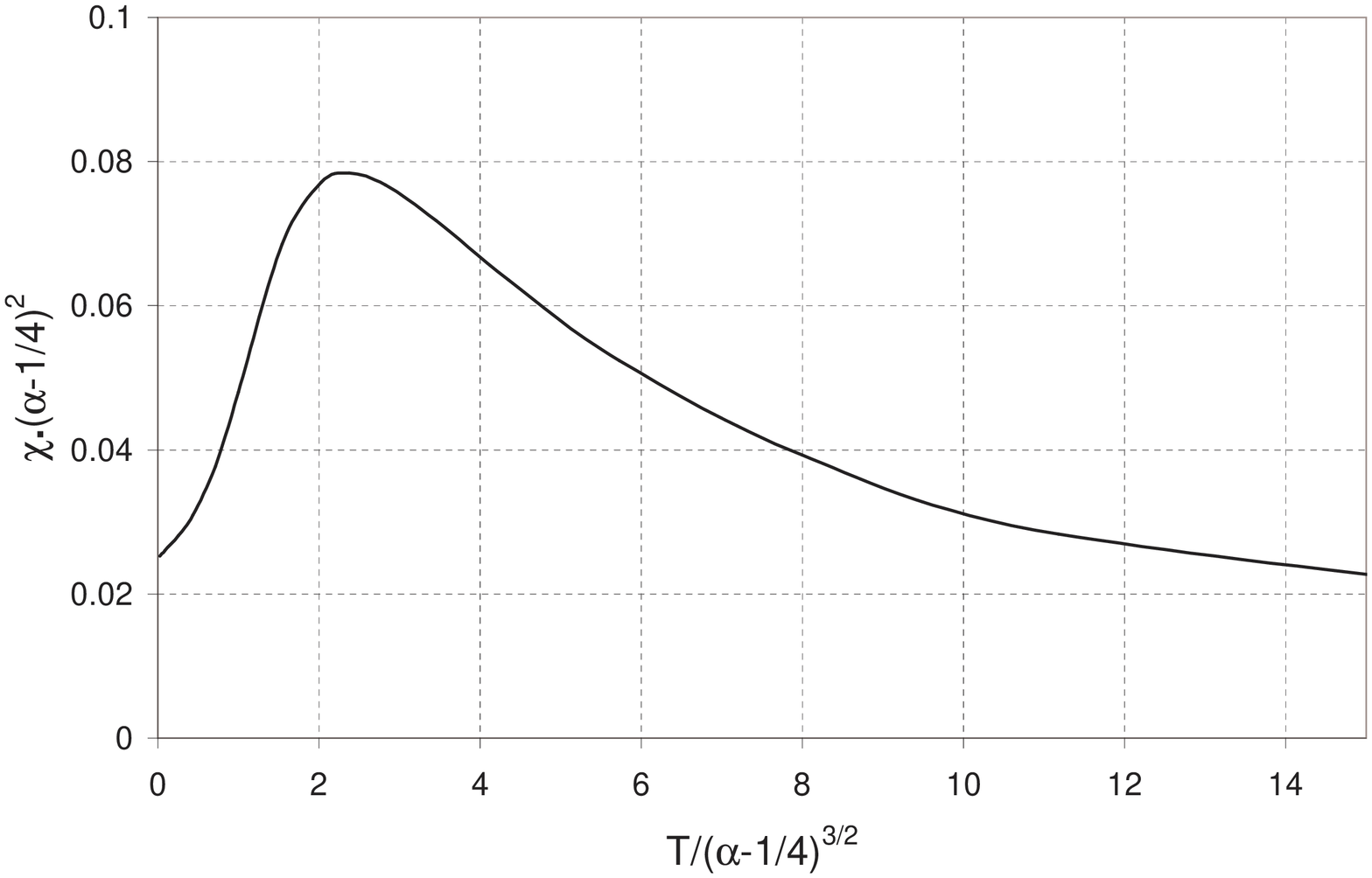}
\caption{Normalized susceptibility $\chi(\alpha -1/4)^2$ as a
function of normalized temperature $T/(\alpha -1/4)^{3/2}$.}
\label{chiTfigure}
\end{figure}

The behavior of the susceptibility in the helical phase is more
interesting. The dependence of the uniform susceptibility
$\widetilde{\chi}$ on the normalized temperature $x=\gamma
^{-3/2}=T/(\alpha -1/4)^{3/2}$ in the helical phase is shown in
Fig.4. The characteristic features of this dependence are the
maximum of $\widetilde{\chi }$ at $x=x_{m}$ and the finite value of
$\widetilde{\chi}$ at $T\to 0$. The latter fact is a classical
effect and can be destroyed by quantum fluctuations. In the quantum
$s=1/2$ F-AF model the ground state is believed to be gapped (though
the gap can be extremely small \cite{Itoi}) and so $\chi\to 0$ at
$T\to 0$. In real materials interchain interactions cause the 3D
spiral LRO and the susceptibility can remain finite at $T=0$.

The obtained dependence $\chi(T)$ is in a qualitative agreement with
those observed in the edge-shared compounds with $\alpha$ close to
$1/4$ ($Li_2CuO_2$ \cite{Mizuno}, $Rb_{2}Cu_{2}Mo_{3}O_{12}$
\cite{Hase}, $Li_{2}CuZrO_{4}$ \cite{Volkova}). As follows from
Eq.(\ref{chiT}) the location of maximum of $\chi(T)$ is at
$T_{m}\sim (\alpha -1/4)^{3/2}$ and $\chi_{m}\sim (\alpha
-1/4)^{-2}$, i.e. with the increase of $\alpha$ the maximum of
$\chi$ shifts to higher temperatures and the magnitude of the
maximum $\chi_{m}$ decreases. These dependencies of $T_{m}$ and
$\chi_{m}$ on the frustration parameter $\alpha$ are also in accord
with the experimental observations \cite{Volkova}.

The dependence $\widetilde{\chi }(x)$ agrees also with the numerical
results for the uniform susceptibility obtained by TMRG and exact
diagonalization methods for the quantum F-AF chain with $s=1/2$
\cite{Lu,Volkova}. The dependence $T_{m}(\alpha )$ is similar to
that obtained by the TMRG calculations in Ref.\cite{Lu}. Thus, the
classical model catches the physics of quantum spin systems in the
helical phase and, therefore, the developed method for the classical
spin model represents a useful tool for the investigation of the
low-temperature behavior of the quantum systems.

\section{Summary}

We have obtained the exact results for the low-temperature
thermodynamics of the classical F-AF model at the frustration
parameter $\alpha =1/4$, where the ground state phase transition
from the ferromagnetic to the helical phase occurs. The main
result relates to the behavior of the zero-field susceptibility
$\chi$ and the correlation length $l_{c}$. It is shown that the
critical exponents of $\chi$ and $l_{c}$ are changed from $2$ to
$4/3$ and from $1$ to $1/3$ correspondingly, when $\alpha \to 1/4$
from the ferromagnetic side. We note that the critical exponents
are the same both for the classical spin model and for the planar
model. In the present paper we have considered a continuum version
of the model (\ref{H14}). However, the low-temperature asymptotes
of $\chi$ and $l_{c}$ of the continuum and lattice models
coincide. In fact, lattice model (\ref{H14}) can be studied on a
base of the transfer matrix method adapted in Ref.[25] to include
the NNN interaction. We have shown [26] that the corresponding
transfer-integral equations are reduced at $T\to 0$ to the
differential equations (\ref{eigen03d}), (\ref{ode2}) with the
same eigenfunctions and eigenvalues, so that the exact
low-temperature asymptotic of the susceptibility coincides with
Eq.(\ref{chi03d}). Besides, the structure factor $S(k)$ for the
lattice model is given by Eq.(\ref{Sk2d}) under substitution of
$2(1-\cos k)$ for $k^2$ in the dominator of Eq.(\ref{Sk2d}). At
the same time, calculations in a frame of path integral method are
essentially simpler and clear in comparison with those in the
transfer-integral approach.

It is interesting to compare the exact expression for the
susceptibility at the transition point with the results found by
approximate methods. One of this methods is the modified spin-wave
theory (MSWT) proposed by Takahashi \cite{Takahashi} to extend the
spin-wave theory to the low-dimensional spin systems without LRO.
Another approximate approach is the expansion of the thermodynamics
of the $n$-vector classical model in powers in $1/n$
\cite{Pesch,Garanin} (usually, up to the first order). Remarkably,
both methods give the true critical exponent $4/3$ for the
susceptibility $\chi =cT^{-4/3}$. However, the numerical coefficient
$c$ differs from the exact one. For example, the MSWT result is
$c=1.19$. The $1/n$ expansion for the classical spin model
(\ref{H14}) gives $c=0.560$ in the zeroth order and $c=0.897$ in the
first order in $1/n$. The comparison of these values of $c$ with the
exact one shows that these approximate methods give a satisfactory
agreement (within $10-20\%$) with the exact coefficient. It is worth
to note that the MSWT gives the exact low-temperature asymptotic of
$\chi=2/3T^{2}$ for the classical ferromagnetic chain ($\alpha=0$)
\cite{Fisher}. Moreover, the MSWT gives the exact result for $\chi$
at $T\to 0$ for the quantum ferromagnetic chain with $s=1/2$ as
well. As was noted in Introduction, the quantum and the classical
ferromagnetic chains have universal low-temperature properties. The
low-temperature susceptibility of the ferromagnetic chain is
described by the scaling function universal for any value of spin
$s$. For the F-AF model at $\alpha =1/4$ there is no rigorous proof
of such universality, though MSWT confirms this hypothesis. If this
universality holds in the case $\alpha =1/4$ then we expect that the
quantum F-AF chain has the same critical exponent of $\chi$ as in
the classical model.

We have also considered the low-temperature thermodynamics in the
vicinity of the transition point. In this case the properties of the
system are governed by the scaling parameter
$\gamma=(\alpha-1/4)/T^{2/3}$. On the ferromagnetic side of the
transition point ($\alpha<1/4$) the susceptibility transforms
smoothly to $\chi\sim(1/4-\alpha)/T^2$ at $\gamma\to -\infty$, which
describes the change in the exponent at $T\to 0$. The susceptibility
in the helical phase for a fixed value of $\alpha\gtrsim 1/4$
behaves as $\chi\sim(\alpha-1/4)^{-2}$ at $\gamma\to\infty$, which
means that the uniform susceptibility remains finite at $T\to 0$.
Besides, we found that the susceptibility in the helical phase has a
maximum at some temperature $T_m\sim(\alpha-1/4)^{-3/2}$. The
presence of maximum of the dependence $\chi(T)$ as well as the
location and the magnitude of this maximum as a function of the
deviation from the transition point $(\alpha-1/4)$ are in agreement
with that observed in several materials described by the quantum
$s=1/2$ version of this model and with the numerical results for the
$s=1/2$ model.

\begin{acknowledgments}
We would like to thank S.-L.Drechsler and J.Sirker for valuable
comments related to this work.
\end{acknowledgments}

\appendix*\section{}

Let us consider a particle moving along a trajectory $\vec{r}(t)$.
At any moment the particle motion can be represented as an instant
rotation around some instantaneous axis of rotation with definite
angular velocity. The radius of the instant rotation $\rho $ is
expressed through the centripetal part $\ddot{\vec{r}}_{c}$ of the
acceleration vector $\ddot{\vec{r}}$:
\begin{equation}
\rho =\frac{\dot{\vec{r}}^{2}}{|\ddot{\vec{r}}_{c}|}  \label{A2}
\end{equation}

The angular velocity is
\begin{equation}
\omega =\frac{|\dot{\vec{r}}|}{\rho }=\frac{|[\dot{\vec{r}}\times \ddot{\vec{%
r}}]|}{\dot{\vec{r}}^{2}}  \label{A4}
\end{equation}
where we took into account that $(\ddot{\vec{r}}_{c}\cdot \dot{\vec{r}})=0$.

Since the rotation takes place in the
$(\dot{\vec{r}},\ddot{\vec{r}})$ plane, the instantaneous axis of
rotation is directed along $[\dot{\vec{r}}\times \ddot{\vec{r}}]$.
Therefore, if we define the `local' coordinate system associated
with the moving particle so that the instant coordinate axes are
directed along the vectors $\dot{\vec{r}}$, $\ddot{\vec{r}}_{c}$ and
$[\dot{\vec{r}}\times \ddot{\vec{r}}]$, then the instant change in
the local coordinate system is expressed by the rotation matrix
\begin{eqnarray}
R &=&\exp \left[ i\left( \vec{\sigma}\cdot \vec{\omega}\right) \right]
\label{A5} \\
\vec{\omega} &=&\frac{[\dot{\vec{r}}\times \ddot{\vec{r}}]}{\dot{\vec{r}}^{2}%
}  \label{A6}
\end{eqnarray}
where
\begin{equation}
\sigma _{x}=\left[
\begin{array}{ccc}
0 & 0 & 0 \\
0 & 0 & i \\
0 & -i & 0%
\end{array}%
\right] ,\qquad \sigma _{y}=\left[
\begin{array}{ccc}
0 & 0 & -i \\
0 & 0 & 0 \\
i & 0 & 0%
\end{array}%
\right] ,\qquad \sigma _{z}=\left[
\begin{array}{ccc}
0 & i & 0 \\
-i & 0 & 0 \\
0 & 0 & 0%
\end{array}%
\right]  \label{A7}
\end{equation}

This implies that if a given fixed vector $\vec{s}$ has components
$\vec{s}_{t0}$ in the local coordinates corresponding to time $t_0$,
then the components of this vector in the local coordinates
corresponding to time $t_1$ can be represented as a chain of
successive rotations:

\begin{equation}
\vec{s}_{t_{1}}=\exp \left[ i\int_{t_{0}}^{t_{1}}\left( \vec{\sigma}\cdot
\vec{\omega}(t)\right) dt\right] \vec{s}_{t_{0}}  \label{A8}
\end{equation}

\end{document}